\documentclass{article}

\usepackage{etex}
\usepackage{hyperref}
\usepackage{authblk}
\usepackage{tabularx}
\usepackage{bussproofs}
\usepackage{proof}
\usepackage{xypic}
\usepackage{fancyhdr}

\usepackage[english]{babel}
\usepackage{url}
\usepackage{algorithm}
\usepackage{algorithmic}
\usepackage{amsmath} 
\usepackage{amssymb}
\usepackage{color}

\newtheorem{definition}{Definition}
\newtheorem{theorem}{Theorem}

\newtheorem{proposition}{Proposition}

\newcommand{\alc}{\ensuremath{\mathcal{ALC}}}
\newcommand{\ialc}{\ensuremath{i\mathcal{ALC}}}
\newcommand{\system}[2]{\ensuremath{\textsf{#1}_{#2}}}
\newcommand{\nominal}[2]{\ensuremath{#1\colon{}\mkern-4mu #2}}

\newcommand{\SEQ}{\ensuremath{\Rightarrow}}

\newcommand{\dland}{\sqcap}        
\newcommand{\dlor}{\sqcup}         
\newcommand{\subs}{\sqsubseteq}    

\def\fCenter{ \SEQ\ }

\newenvironment{dedsystem}
{
 \newcolumntype{Y}{>{\centering\arraybackslash}X}
 \setlength{\extrarowheight}{2.5ex}
 \tabularx{\textwidth}{YcY}}
{\endtabularx}

\title{An Intuitionisticaly based Description Logic}

\author[1]{Alexandre Rademaker}
\author[2]{Edward Hermann Haeusler}
\affil[1]{IBM Research and FGV/EMAp, \texttt{arademaker@br.ibm.com}}
\affil[2]{Dep. Inform\'atica, PUC-Rio, \texttt{hermann@inf.puc-rio.br}}
\date{}

\begin{document}

\maketitle

\begin{abstract}
  This article presents $\ialc$, an intuitionistic version of the
  classical description logic \alc, based on the framework for
  constructive modal logics presented by Simpson~\cite{simpson95} and
  related to description languages, via hybrid logics, by
  dePaiva~\cite{depaiva2003}.  This article corrects and extends the
  presentation of \ialc~ appearing in \cite{PHR:2010}. It points out
  the difference between \ialc~ and the intuitionistic hybrid logic
  presented in \cite{depaiva2003}. Completeness and soundness proofs
  are provided. A brief discussion on the computational complexity of
  \ialc~ provability is taken.  It is worth mentioning that \ialc~ is
  used to formalize legal knowledge \cite{HPR:2010a,HPR:2010ab,Jurix,
    HPR:2011}, and in fact, was specifically designed to this goal.
\end{abstract}

\section{Intuitionistic ALC}\label{sec:ialc}

The \ialc\ logic is based on the framework for intuitionistic modal
logic IK proposed in \cite{simpson95,Fisher1984,plotkin1986}. These
modal logics arise from interpreting the usual possible worlds
definitions in an intuitionistic meta-theory. As we will see in the
following paragraphs, ideas from \cite{brauner-depaiva2006} were also
used, where the framework IHL, for \emph{intuitionistic hybrid
  logics}, is introduced. \ialc\ concepts are described as:

\[
C, D ::= A \mid \bot \mid \top
	   \mid \neg C
           \mid C \dland D 
           \mid C \dlor  D 
           \mid C \subs D 
           \mid \exists R.C 
           \mid \forall R.C 
\]
where $C,D$ stands for concepts, $A$ for an atomic concept, $R$ for an
atomic role.
We could have used distinct symbols for subsumption of concepts and
the subsumption concept constructor but this would blow-up the
calculus presentation. This syntax is more general than standard \alc\
since it includes subsumption $\subs$ as a concept-forming
operator. We have no use for nested subsumptions, but they do make the
system easier to define, so we keep the general rules. Negation could
be defined via subsumption, that is, $\neg C=C\subs\bot$, but we find
it convenient to keep it in the language.  The constant $\top$ could
also be omitted since it can be represented as $\neg\bot$.

A constructive interpretation of \ialc\ is a structure $\mathcal{I}$
consisting of a non-empty set $\Delta^{\mathcal{I}}$ of entities in
which each entity represents a partially defined individual; a
refinement pre-ordering $\preceq^{\cal I}$ on $\Delta^{\cal I}$, i.e.,
a reflexive and transitive relation; and an interpretation function
$\cdot^{\cal I}$ mapping each role name $R$ to a binary relation
$R^{\cal I}\subseteq\Delta^{\cal I} \times \Delta^{\cal I}$ and atomic
concept $A$ to a set $A^{\cal I}\subseteq\Delta^{\cal I}$ which is
closed under refinement, i.e., $x \in A^{\cal I}$ and $x\preceq^{\cal
  I}y$ implies $y\in A^{\cal I}$. The interpretation $\cal I$ is
lifted from atomic concepts to arbitrary concepts via:
\[
\begin{array}{rl}
\top^{\cal I}  & =_{df} \Delta^{\cal I} \\
\bot^{\cal I}  & =_{df} \emptyset \\
(\neg C)^{\cal I } & =_{df} \{x \mid \forall y \in \Delta^{\cal I} . x \preceq y\Rightarrow  y \not\in C^{\cal I}  \} \\
(C\dland  D)^{\cal I } & =_{df} C^{\cal I}\cap  D^{\cal I} \\
(C \dlor D)^{\cal I } & =_{df} C^{\cal I} \cup D^{\cal I } \\
(C \subs D)^{\cal I } & =_{df } \{x \mid \forall y \in  \Delta^{\cal I} . 
(x \preceq y  \text{ and }  y \in C^{\cal I }) \Rightarrow y \in D^{\cal I}  \} \\
(\exists R.C)^{\cal I} & =_{df} \{ x \mid \exists y \in \Delta^{\cal I} . (x,y) \in  R^{\cal I} \mbox{ and } y \in C^{\cal I} \} \\
(\forall R.C)^{\cal I}  & =_{df} \{ x \mid \forall y \in  \Delta^{\cal I} . x \preceq y 
  \Rightarrow \forall z \in \Delta^{\cal I}. (y, z ) \in R^{\cal I} \Rightarrow z\in C^{\cal I}  \} \\
\end{array}
\]

Following the semantics of IK, the structures $\cal I$ are models for
\ialc\ if they satisfy two frame conditions:
\begin{description}
\item[F1] if $w \leq w'$ and $wRv$ then $\exists v' . w'Rv'$ and $v \leq v'$
\item[F2] if $v \leq v'$ and $wRv$ then $\exists w' . w'Rv'$ and $w \leq w'$
\end{description}

The above conditions are diagrammatically expressed as:
\[
\begin{array}{ccc}
\xymatrix{
w^{\prime}\ar@{.>}[r]^{R} \ar@{}[dr]|{(F1)} & v^{\prime} \\
w \ar[r]^{R} \ar[u]^{\le} & v\ar@{.>}[u]_{\le}
}
& \mbox{and} &
\xymatrix{
w^{\prime}\ar@{.>}[r]^{R} \ar@{}[dr]|{(F2)} & v^{\prime} \\
w \ar[r]^{R} \ar@{.>}[u]^{\le} & v\ar[u]_{\le}
}
\end{array}
\]

Our setting simplifies \cite{mendler-scheele}, since $i\mathcal{ALC}$
satisfies (like classical $\mathcal{ALC}$) $\exists R.\bot=\bot$ and
$\exists R.(C\dlor D) = \exists R.C\dlor \exists R.D$.

Building up from the Simpson's constructive modal logics (called here
IML), in \cite{brauner-depaiva2006}, it is introduced
\emph{intuitionistic hybrid logics}, denoted by \textsf{IHL}. Hybrid
logics add to usual modal logics a new kind of propositional symbols,
the \emph{nominals}, and also the so-called \emph{satisfaction
  operators}. A nominal is assumed to be true at exactly one world, so
a nominal can be considered the name of a world. If $x$ is a nominal
and $X$ is an arbitrary formula, then a new formula \nominal{x}{X}
called a satisfaction statement can be formed. The satisfaction
statement \nominal{x}{X} expresses that the formula $X$ is true at one
particular world, namely the world denoted by $x$. In hindsight one
can see that IML shares with hybrid formalisms the idea of making the
possible-world semantics part of the deductive system. While IML makes
the relationship between worlds (e.g., $xRy$) part of the deductive
system, IHL goes one step further and sees the worlds themselves $x,y$
as part of the deductive system, (as they are now nominals) and the
satisfaction relation itself as part of the deductive system, as it is
now a syntactic operator, with modality-like properties. In contrast
with the above mentioned approaches, ours assign a truth values to
some formulas, also called assertions, they are not concepts as in
\cite{brauner-depaiva2006}, for example.  Below we define the syntax
of general assertions ($A$) and nominal assertions ($N$) for ABOX
reasoning in \ialc. Formulas ($F$) also includes subsumption of
concepts interpreted as propositional statements.
\[
N ::=  x\colon C \mid x\colon N
\hspace{2cm}
A ::=  N \mid x R y \hspace{2cm}  F ::= A \mid C \subs C 
\]
where $x$ and $y$ are nominals, $R$ is a role symbol and $C$ is a
concept. In particular, this allows $x\colon (y\colon C)$, which is a
perfectly valid nominal assertion.

\begin{definition}[outer nominal]\label{def:outer}
  In a nominal assertion $x\colon\gamma$, $x$ is said to be the outer
  nominal of this assertion. That is, in an assertion of the form
  $x\colon(y\colon\gamma)$, $x$ is the outer nominal.
\end{definition}

We write $\mathcal{I},w \models C$ to abbreviate $w \in C^\mathcal{I}$
which means that entity $w$ satisfies concept $C$ in the
interpretation $\cal I$\footnote{In IHL, this $w$ is a world and this
  satisfaction relation is possible world semantics}. Further, $\cal
I$ is a model of $C$, written $\mathcal{I} \models C$ iff $\forall w
\in {\cal I} . {\cal I}, w \models C$. Finally, $\models C$ means
$\forall {\cal I}.{\cal I} \models C$. All previous notions are
extended to sets $\Phi$ of concepts in the usual universal
fashion. Given the hybrid satisfaction statements, the interpretation
and semantic satisfaction relation are extended in the expected
way. The statement $\mathcal{I},w\models x\colon C$ holds, if and only
if, $\forall z_{x}\succeq^{\cal I} x\ .\ \mathcal{I},z_{x} \models C$.
In a similar fashion, $\mathcal{I}, w \models xRy$ holds ,if and only
if, $\forall z_{x}\succeq x.\forall z_{y}\succeq
y.(x_{x}^{\mathcal{I}},z_{y}^{\mathcal{I}}) \in R^{\mathcal{I}}$. That
is, the evaluation of the hybrid formulas does not take into account
only the world $w$, but it has to be monotonically preserved. It can
be observed that for every $w^{\prime}$, if $x^{\mathcal{I}}\preceq
w^{\prime}$ and $\mathcal{I},x^{\prime}\models \alpha$, then
$\mathcal{I},w^{\prime}\models \alpha$ is a property holding on this
satisfaction relation.

In common reasoning tasks the interpretation $\cal I$ and the entity
$w$ in a verification goal such as ${\cal I}, w \models\delta$ are not
given directly but are themselves axiomatized by sets of concepts and
formulas. Usually we have a set $\Theta$~\footnote{Here we consider
  only acycled TBox with $\subs$ and $\equiv$.} of formulas and the
set $\Gamma$ of concepts. Accordingly:

\begin{definition}\label{models}
We write $\Theta,\Gamma\models\delta$ if it is the case
that:
\begin{multline}
\forall {\cal I}.((\forall x \in \Delta^{\cal I}.({\cal I}, x \models\Theta))  \\ 
\Rightarrow \forall(Nom(\Gamma,\delta)).\forall \vec{z}\succeq Nom(\Gamma,\delta).( {\cal I}, \vec{z} \models \Gamma 
\Rightarrow {\cal I},\vec{z} \models \delta)
\end{multline}
where $\vec{z}$ denotes a vector of variables $z_1,\ldots,z_k$ and
$Nom(\Gamma,\delta)$ is the vector of all outer nominals occurring in
each nominal assertion of $\Gamma\cup\{\delta\}$. $x$ is the only
outer nominal of a nominal assertion $\{x\colon \gamma\}$, while a
(pure) concept $\gamma$ has no outer nominal.
\end{definition}

A Hilbert calculus for \ialc\ is provided following
\cite{plotkin1986,simpson95,Fisher1984}. It consists of all axioms of
intuitionistic propositional logic plus the axioms and rules displayed
in Figure~\ref{fig:axioms}. The Hilbert calculus implements
TBox-reasoning. That is, it decides the semantical relationship
$\Theta, \emptyset \models C$. $\Theta$ has only formulas as members.

\begin{figure}[htbp]
\begin{align*}
0. & \quad \mbox{all substitution instances of theorems of IPL}      \\
1. & \quad \forall R.(C\subs D)\subs (\forall R.C\subs \forall R.D)  \\
2. & \quad \exists R.(C\subs D)\subs (\exists R.C\subs \exists R.D)  \\
3. & \quad \exists R.(C\dlor D)\subs (\exists R.C\dlor \exists R.D)  \\
4. & \quad \exists R.\bot\subs \bot                                  \\
5. & \quad (\exists R.C\subs \forall R.C)\subs \forall R.(C\subs D)  \\
\mathsf{MP}  & \quad \mbox{If $C$ and $C\subs D$ are theorems, $D$ is a theorem too.} \\
\mathsf{Nec} & \quad \mbox{If $C$ is a theorem then $\forall R.C$ is a theorem too.} 
\end{align*}
\caption{The \ialc\ axiomatization}\label{fig:axioms}
\end{figure}

A Sequent Calculus for \ialc\ is also provided. The logical rules of
the Sequent Calculus for \ialc\ are presented in
Figure~\ref{fig:sc-ialc}.~\footnote{The reader may want to read Proof
  Theory books, for example,
  \cite{takeuti2013,buss1998,negri2008,girard1989}.} The structural
rules and the cut rule are omitted but they are as usual. The $\delta$
stands for concepts or assertions ($x\colon C$ or $xRy$), $\alpha$ and
$\beta$ for concept and $R$ for role. $\Delta$ is a set of
formulas. In rules p-$\exists$ and p-$\forall$, the syntax $\forall
R.\Delta$ means $\{\forall R.\alpha \mid \alpha\in concepts(\Delta)
\}$,
that is, all concepts in $\Delta$ are universal quantified with the
same role. The assertions in $\Delta$ are kept unmodified. In the same
way, in rule p-N the addition of the nominal is made only in the
concepts of $\Delta$ (and in $\delta$ if that is a concept) keeping
the assertions unmodified.

The propositional connectives ($\dland,\dlor,\subs$) rules are as
usual, the rule $\dlor_2$-r is omitted. The rules are presented
without nominals but for each of these rules there is a counterpart
with nominals. For example, the rule $\subs$-r has one similar:
\begin{prooftree}
\Axiom$ \Delta, x\colon\alpha \fCenter x\colon\beta $ \RightLabel{ n-$\subs$-r }
\UnaryInf$ \Delta \fCenter x\colon(\alpha\subs\beta)$
\end{prooftree}

The main modification comes for the modal rules, which are now role
quantification rules. We must keep the intuitionistic constraints for
modal operators. Rule~$\exists$-l has the usual condition that $y$ is
not in the conclusion. Concerning the usual condition on the
$\forall$-r rule, it is not the case in this system, for the
interpretation of the a nominal assertion in a sequent is already
implicitly universal (Definition~\ref{models}).

\begin{figure}[htbp]
\begin{center}
\begin{dedsystem}
\hline
\AxiomC{}
\UnaryInf$\Delta, \delta \fCenter \delta $
\DisplayProof & & 
\AxiomC{}
\UnaryInf$\Delta, x\colon\bot \fCenter \delta $
\DisplayProof  \\
\Axiom$ \Delta, xRy \fCenter y\colon\alpha$  
\RightLabel{$\forall$-r}
\UnaryInf$ \Delta \fCenter x\colon\forall R.\alpha$
\DisplayProof & &
\Axiom$ \Delta, x\colon\forall R.\alpha, y\colon\alpha, xRy \fCenter \delta$ 
\RightLabel{$\forall$-l}
\UnaryInf$\Delta, x\colon\forall R.\alpha, xRy \fCenter \delta$
\DisplayProof \\ 
\Axiom$\Delta \fCenter xRy $
\Axiom$\Delta \fCenter y\colon\alpha$ 
\RightLabel{$\exists$-r}
\BinaryInf$\Delta \fCenter x\colon\exists R.\alpha$
\DisplayProof & & 
  \Axiom$ \Delta, xRy, y\colon\alpha \fCenter \delta $ 
\RightLabel{$\exists$-l}
\UnaryInf$  \Delta, x\colon\exists R.\alpha \fCenter \delta $
\DisplayProof \\
    \Axiom$ \Delta, \alpha \fCenter \beta $ \RightLabel{ $\subs$-r }
 \UnaryInf$ \Delta \fCenter \alpha \subs \beta $
 \DisplayProof  & &
     \Axiom$ \Delta_1 \fCenter \alpha $
     \Axiom$ \Delta_2, \beta \fCenter \delta $ \RightLabel{ $\subs$-l }
 \BinaryInf$ \Delta_1, \Delta_2, \alpha \subs \beta \fCenter \delta $
 \DisplayProof \\
     \Axiom$ \Delta \fCenter \alpha $
     \Axiom$ \Delta \fCenter \beta  $ \RightLabel{ $\dland$-r }
 \BinaryInf$ \Delta \fCenter \alpha \sqcap \beta  $
 \DisplayProof & & 
    \Axiom$  \Delta, \alpha, \beta \fCenter \delta $ \RightLabel{ $\dland$-l }
 \UnaryInf$  \Delta, \alpha \sqcap \beta \fCenter \delta $
 \DisplayProof \\
    \Axiom$  \Delta \fCenter \alpha $ \RightLabel{ $\dlor_1$-r }
 \UnaryInf$ \Delta \fCenter \alpha\sqcup\beta $
 \DisplayProof & & 
     \Axiom$ \Delta, \alpha \fCenter \delta $
     \Axiom$ \Delta, \beta  \fCenter \delta  $ \RightLabel{ $\dlor$-l }
 \BinaryInf$ \Delta, \alpha\sqcup\beta \fCenter \delta  $
 \DisplayProof \\
   \Axiom$ \Delta, \alpha \fCenter \beta $ \RightLabel{ p-$\exists$ }
\UnaryInf$ \forall R.\Delta, \exists R.\alpha \fCenter \exists R.\beta $
\DisplayProof && 
   \Axiom$ \Delta \fCenter \alpha $\RightLabel{ p-$\forall$ } 
\UnaryInf$ \forall R.\Delta \fCenter \forall R.\alpha $
\DisplayProof \\ 
   \Axiom$ \Delta \fCenter \delta $ \RightLabel{p-N}
\UnaryInf$ x\colon\Delta \fCenter x\colon\delta $
\DisplayProof & \\[3ex]\hline
\end{dedsystem}
\caption{The System \system{SC}{iALC}: logical rules}\label{fig:sc-ialc}
\end{center}
\end{figure}

\begin{theorem} 
  The sequent calculus described in Fig.~\ref{fig:sc-ialc} is sound
  and complete for TBox reasoning, that is $\Theta,\emptyset\models C$
  if and only if $\Theta\Rightarrow C$ is derivable with the rules of
  Figure~\ref{fig:sc-ialc}.
\end{theorem}

The completeness of our system is proved relative to the
axiomatization of \ialc, shown in Figure~\ref{fig:axioms}. The proof
is presented in Section~\ref{sec:completeness}.

The soundness of the system is proved directly from the semantics of
\ialc\ including the ABOX, that is, including nominals. The semantics
of a sequent is defined by the satisfaction relation, as shown in
Definition~\ref{models}. The sequent $\Theta,\Gamma \SEQ\ \delta$ is
valid if and only if $\Theta, \Gamma \models \gamma$. Soundness is
proved by showing that each sequent rule preserves the validity of the
sequent and that the initial sequent is valid. This proof is presented
in Section~\ref{sec:soundness}.

We note that although we have here fixed some inaccuracies in the
presentation of the \ialc\ semantics in \cite{PHR:2010}, the system
presented here is basically the same, excepted that here the
propositional rules are presented without nominals. Given that, the
soundness of the system proved in \cite{PHR:2010} can be still
considered valid without further problems. Note also that the proof of
soundness provides in Section~\ref{sec:soundness} is regarded the full
language of \ialc. It considers nominals and assertion on nominals
relationship, that is it concerns ABOX and TBOX. The proof of
completeness is for the TBOX only. A proof of completeness for ABOX
can be done by the method of canonical models. For the purposes of
this article, we choose to show the relative completeness proof with
the sake of showing a simpler proof concerning TBOX.

\section{The completeness of \system{SC}{iALC} system}
\label{sec:completeness}

We show the relative completeness of \system{SC}{iALC} regarding the
axiomatic presentation of \ialc\ presented in
Figure~\ref{fig:axioms}. To prove the completeness of
\system{SC}{iALC} it is sufficient to derive in \system{SC}{iALC} the
axioms 1--5 of \ialc. It is clear that all substitution instances of
IPL theorems can also be proved in \system{SC}{iALC} using only
propositional rules. The MP rule is a derived rule from the
\system{SC}{iALC} using the cut rule. The Nec rule is the p-$\forall$
rule in the system with $\Delta$ empty. In the first two proofs below
do not use nominals for given better intuition of the reader about the
use of rules with and without nominals.

\vspace{.5cm}

\noindent Axiom 1:
\begin{prooftree}
\def\fCenter{ \SEQ\ }
\Axiom$ \alpha \fCenter \alpha $
\Axiom$ \beta \fCenter \beta $ \RightLabel{$\subs$-l}
\BinaryInf$ \alpha \subs \beta, \alpha \fCenter \beta $ \RightLabel{p-$\exists$}
\UnaryInf$ \forall R.(\alpha \subs \beta), \exists R.\alpha \fCenter  \exists R.\beta $ \RightLabel{$\subs$-r}
\UnaryInf$ \forall R.(\alpha \subs \beta) \fCenter \exists R.\alpha \subs \exists R.\beta $ 
\end{prooftree}

\noindent Axiom 2:
\begin{prooftree}
\def\fCenter{ \SEQ\ }
\Axiom$ \alpha \fCenter \alpha $
\Axiom$ \beta \fCenter \beta $ \RightLabel{$\subs$-l}
\BinaryInf$ \alpha \subs \beta, \alpha \fCenter \beta $ \RightLabel{p-$\forall$}
\UnaryInf$ \forall R.(\alpha \subs \beta), \forall R.\alpha \fCenter  \forall R.\beta $ \RightLabel{$\subs$-r}
\UnaryInf$ \forall R.(\alpha \subs \beta) \fCenter \forall R.\alpha \subs \forall R.\beta $ 
\end{prooftree}

\noindent Axiom 3:
\begin{prooftree}
\def\fCenter{ \SEQ\ }
\Axiom$ xRy, y:\bot \fCenter x:\bot $ \RightLabel{$\exists$-l}
\UnaryInf$ x:\exists R.\bot \fCenter x:\bot $ \RightLabel{$\subs$-r}
\UnaryInf$ \fCenter x:(\exists R.\bot \subs \bot) $
\end{prooftree}

\noindent Axiom 4:
\begin{prooftree}\small
\def\fCenter{ \SEQ\ }
   \Axiom$ x:\exists R.\alpha \fCenter x:\exists R.\alpha $ 
\RightLabel{$\dlor_1$-r} \UnaryInf$ x:\exists R.\alpha \fCenter x:(\exists R.\alpha \dlor \exists R.\beta) $ 

   \Axiom$ x:\exists R.\beta \fCenter x:\exists R.\beta $ 
\RightLabel{$\dlor_2$-r} \UnaryInf$ x:\exists R.\beta \fCenter x:(\exists R.\alpha \dlor \exists R.\beta) $ 
\RightLabel{$\dlor$-l}   \BinaryInf$ x:\exists R.(\alpha \dlor \beta) \fCenter x:(\exists R.\alpha \dlor \exists R.\beta) $ 
\end{prooftree}

\noindent Axiom 5:
\begin{prooftree}\small
\def\fCenter{ \SEQ\ }
\AxiomC{$ xRy, y:\alpha \fCenter y:\alpha $}
\AxiomC{$ xRy, y:\alpha \fCenter xRy $}
\RightLabel{$\exists$-r} 
\BinaryInfC{$ xRy, y:\alpha \fCenter {\color{red} x:}\exists R.\alpha $}
\AxiomC{$ xRy, y:\alpha, y:\beta, \forall R.\beta \fCenter y:\beta $}
\RightLabel{$\forall$-l} \UnaryInfC{$ xRy, y:\alpha, {\color{red} x:}\forall R.\beta \fCenter y:\beta$}
\RightLabel{$\subs$-l}  \BinaryInfC{$ x:(\exists R.\alpha \subs \forall R.\beta), xRy, y:\alpha \fCenter y:\beta$}
\RightLabel{$\forall$-r} \UnaryInfC{$ x:(\exists R.\alpha \subs \forall R.\beta), xRy \fCenter y:(\alpha \subs \beta)$} 
\RightLabel{$\forall$-r} \UnaryInfC{$ x:(\exists R.\alpha \subs \forall R.\beta) \fCenter x:\forall R.(\alpha \subs \beta)$}
\RightLabel{$\subs$-r}   \UnaryInfC{$ \fCenter x:[(\exists R.\alpha \subs \forall R.\beta) \subs \forall R.(\alpha \subs \beta)]$}
\end{prooftree}

\section{Soundness of \system{SC}{iALC} system}
\label{sec:soundness}

In this section we prove that.

\begin{proposition} 
  If $\Theta,\Gamma \SEQ\ \delta$ is provable in \system{SC}{iALC}
  then $\Theta, \Gamma \models \gamma$.
\end{proposition}

{\em Proof}: We prove that each sequent rule preserves the validity of
the sequent and that the initial sequents are valid. The definition of
a valid sequent ($\Theta,\Gamma\models\gamma$) is presented in
Definition~\ref{models}.

The validity of the axioms is trivial. We first observe that any
application of the rules $\subs$-r, $\subs$-l,$\dland$-r,$\dland$-l,
$\dlor_1$-r,$\dlor_2$-r, $\dlor$-l of \system{SC}{iALC} where the
sequents do not have any nominal, neither in $\Theta$ nor in $\Gamma$,
is sound regarded intuitionistic propositional logic kripke semantics,
to which the validity definition above collapses whenever there is no
nominal in the sequents. Thus, in this proof we concentrate in the
case where there are nominals. We first observe that the nominal
version of $\subs$-r, the validity of the premises includes
\[
\forall(Nom(\Gamma,\delta)).\forall \vec{z}\succeq Nom(\Gamma,\delta).( {\cal I}, \vec{z} \models 
\Gamma \Rightarrow {\cal I},\vec{z} \models \delta)
\]
This means that $\Gamma$ holds in any worlds $\vec{z}\succeq\vec{x}$
for the vector $\vec{x}$ of nominals occurring in $\Gamma$. This
includes the outer nominal $x_i$ in $\delta$ (if any). In this case
the semantics of $\subs$ is preserved, since $\vec{z}$ includes
$z_i\succeq x_i$. With the sake of a more detailed analysis, we
consider the following instance:

\begin{prooftree}
\def\fCenter{ \SEQ\ }
\Axiom$ x:\alpha_1, y:\alpha_2 \fCenter x:\beta $ \RightLabel{ $\subs$-r }
\UnaryInf$ \alpha_1 \fCenter x:\alpha_2 \subs \beta $
\end{prooftree}

Consider an \ialc~structure ${\cal I}=\langle{\cal U},\preceq,R^{\cal
  I}\ldots,C^{\cal I}\rangle$ In this case, for any ${\cal I}$ and any
$z_1,z_2\in{\cal U}^{\cal I}$ if $z_1\succeq x^{\cal I}$, $z_1\succeq
y^{\cal I}$, such that, ${\cal I}, z_{i} \models \alpha_1$ and ${\cal
  I}, z_{i} \models \alpha_2$, we have that ${\cal I},z_i \models
x:\beta$, since the premise is valid, by hypothesis. In this case, by
the semantics of $\subs$ we have ${\cal I}, z_i\models
x:\alpha_1\subs\beta$. The conclusion of the rule is valid too.

The argument shown above for the $\subs$-r rule is analogous for the
nominal versions of $\subs$-r, $\subs$-l,$\dland$-r,$\dland$-l,
$\dlor_1$-r,$\dlor_2$-r, $\dlor$-l. Consider the rule $\forall$-r. 

\begin{prooftree}
\def\fCenter{ \SEQ\ }
\Axiom$ \Delta, xRy \fCenter y\colon\alpha$  
\RightLabel{$\forall$-r}
\UnaryInf$ \Delta \fCenter x\colon\forall R.\alpha$
\end{prooftree}

Since the premise is valid we have that if $\forall z_{x}\succeq
x^{\cal I}$, $\forall z_{y}\succeq y^{\cal I}$, $(z_{x},z_{y})\in
R^{\cal I}$ then $\forall z_{y}\succeq y^{\cal I}.{\cal
  I},z_{y}\models \gamma$. This entails that $x^{\cal I}\in (\forall
R.\gamma)^{\cal I}$, for $x^{\cal I}\succeq x^{\cal I}$. We observe
that by the restriction on the rule application, $y$ does not occur in
$\Delta$, it only occurs in $xRy$ and $y:\alpha$. The truth of these
formulas are subsumed by $\forall R.\gamma$. The conclusion does not
need to consider them any more. The conclusion is valid too. Another
way to see its soundness is to prove that if $xRy\Rightarrow y:\alpha$
is valid, then so is $\Rightarrow x:\forall R.\alpha$. This can be
show by the following reasoning:
\[
\forall x^{\cal I}\forall y^{\cal I}\forall z_{x}\forall
z_{y}(z_{x}\succeq x^{\cal I}\rightarrow (z_{y}\succeq y^{\cal
  I}\rightarrow ((z_{x},z_{y})\in R^{\cal I}\rightarrow {\cal
  I},z_y\models y:\alpha)))
\]
that is the same as:
\[
\forall x^{\cal I}\forall y^{\cal I}\forall z_{x}\forall
z_{y}(z_{x}\succeq x^{\cal I}\rightarrow (z_{y}\succeq y^{\cal
  I}\rightarrow ((z_{x},z_{y})\in R^{\cal I}\rightarrow {\cal
  I},y^{\cal I}\models\alpha)))
\]
Using the fact that $\forall y^{\cal I}(y^{\cal I}\succeq y^{\cal I})$,  we obtain:
\[
\forall x^{\cal I}\forall z_{x}(z_{x}\succeq x^{\cal I}\rightarrow
\forall y^{\cal I}((z_{x},y^{\cal I})\in R^{\cal I}\rightarrow {\cal
  I},y^{\cal I}\models\alpha))
\]
The above condition states that $\Rightarrow x:\forall R.\alpha$ is valid. 
\[
\forall x^{\cal I}\forall y^{\cal I}\forall z_{x}\forall
z_{y}(z_{x}\succeq x^{\cal I}\rightarrow (z_{y}\succeq y^{\cal
  I}\rightarrow ((z_{x},z_{y})\in R^{\cal I}\rightarrow {\cal
  I},z_y\models y:\alpha)))
\]

Consider the rule $\forall$-l:

\begin{prooftree}
\def\fCenter{ \SEQ\ }
\Axiom$ \Delta, x\colon\forall R.\alpha, y\colon\alpha, xRy \fCenter \delta$ 
\RightLabel{$\forall$-l}
\UnaryInf$\Delta, x\colon\forall R.\alpha, xRy \fCenter \delta$
\end{prooftree}

As in the $\forall$-r case, we analyze the simplest validity
preservation: if $x:\forall R.\alpha\land xRy$ is valid, then so is
$x:\forall R.\alpha\land y:\alpha\land xRy$. The first condition is:

\begin{multline}
\forall x^{\cal I} \forall y^{\cal I} \forall z_x 
(z_{x}\succeq x^{\cal I} \rightarrow \forall z_{y}(z_{y}\succeq y^{\cal I} \rightarrow \\
(({\cal I},z_y\models x:\forall R.\alpha)\land ({\cal I},z_y\models
x:\forall R.\alpha)\land ((z_{x},z_{y})\in R^{\cal I}) \rightarrow \\
({\cal I},z_y\models y:\alpha)\land ({\cal I},z_x\models y:\alpha))))
\end{multline}

Using $z_y=y^{\cal I}$, eliminating $z_x$ from the term, and, using
the fact that ${\cal I},z_y\models y:\alpha$ is valid, iff, ${\cal
  I},y^{\cal I}\models\alpha$ , we obtain

\begin{multline}
\forall x^{\cal I}\forall y^{\cal I}\forall z_{x}(z_{x}\succeq x^{\cal
  I}\rightarrow \forall z_{y}(z_{y}\succeq y^{\cal I}\rightarrow \\
(({\cal I},z_y\models x:\forall R.\alpha)\land ({\cal I},z_y\models
x:\forall R.\alpha)\land ((z_{x},z_{y})\in R^{\cal I})\rightarrow
({\cal I},y\models\alpha))))
\end{multline}

Consider the semantics of $\exists R.\alpha$:

\[
(\exists R.\alpha)^{\cal I}=_{df}\{ x \mid \exists y \in {\cal
  U}^{\cal I} . (x,y) \in R^{\cal I} \mbox{ and } y \in \alpha^{\cal
  I} \}
\]
and the following rule:

\begin{prooftree}
\def\fCenter{ \SEQ\ }
\Axiom$\Delta \fCenter xRy $
\Axiom$\Delta \fCenter y\colon\alpha$ 
\RightLabel{$\exists$-r}
\BinaryInf$\Delta \fCenter x\colon\exists R.\alpha$
\end{prooftree}

We can see that the premises of the rule entails the conclusion.  The
premises correspond to the following conditions:
\[
\forall x^{\cal I} \forall y^{\cal I}\forall z_{x}(z_{x}\succeq
x^{\cal I}\rightarrow \forall z_{y}(z_{y}\succeq y^{\cal I}\rightarrow
((z_{x},z_{y})\in R^{\cal I})))
\]
and
\[
\forall y^{\cal I}\forall z_{y}(z_{y}\succeq y^{\cal I}\rightarrow
(({\cal I},z_y\models y:\alpha)))
\]
Instantiating in both conditions $z_{y}=y^{\cal I}$ and $z_{x}=x^{\cal
  I}$, this yields $(x^{\cal I},y^{\cal I})\in R^{\cal I}$, such that
${\cal I},y^{\cal I}\models \alpha$, so ${\cal I}, z_{x}\models
x^{\cal I}:\exists R.\alpha$. Thus, $\exists$-r is sound. The
soundness of $\exists$-l is analogous to $\forall$-l.

Finally, it is worth noting that, for each rule, we can derive the
soundness of its non-nominal version from the proof of soundness of
its nominal version. For instance, the soundness of the nominal
version of rule~$\dlor$-l depends on the diamond conditions F1 and
F2. The soundness of its non-nomimal version, is a consequence of the
soundness of the nominal version.

The rules below have their soundness proved as a consequence of the
following reasonings in first-order intuitionistic logic that are used
for deriving the semantics of the conclusions from the semantics of
the premises: 
\begin{description}
\item[(p-$\exists$)] $\forall x(A(x)\land B(x)\rightarrow C(x))\models
  \forall x A(x)\land\exists xB(x)\rightarrow \exists xC(x)$;

\item[(p-$\forall$)] $(A(x)\models B(x))$ implies $\forall
  y(R(y,x)\rightarrow A(x))\models\forall y(R(y,x)\rightarrow B(x))$;

\item[(p-N)] if $A\models B$ then for every Kripke model ${\cal
    I}$ and world $x^{\cal I}$, if ${\cal I},x^{\cal I}\models A$ then
  ${\cal I}, x^{\cal I}\models B$.
\end{description}

\vspace{0.5cm}
\noindent \begin{tabularx}{\textwidth}{ccc}
   \Axiom$ \Delta, \alpha \fCenter \beta $ \RightLabel{p-$\exists$}
\UnaryInf$ \forall R.\Delta, \exists R.\alpha \fCenter \exists R.\beta $
\DisplayProof & 
   \Axiom$ \Delta \fCenter \alpha $\RightLabel{ p-$\forall$} 
\UnaryInf$ \forall R.\Delta \fCenter \forall R.\alpha $
\DisplayProof & 
   \Axiom$ \Delta \fCenter \delta $ \RightLabel{p-N}
\UnaryInf$ x\colon\Delta \fCenter x\colon\delta $
\DisplayProof \\
\end{tabularx}

\bibliographystyle{plain}
\bibliography{iALC}

\end{document}